\documentstyle[12pt] {article}
\def\zid{1\kern-0.36em\llap~1}

\newcommand{\beq}{\begin{equation}}
\newcommand{\ber}{\begin{eqnarray}}
\newcommand{\eeq}{\end{equation}}
\newcommand{\eer}{\end{eqnarray}}

\begin{document}

\begin{titlepage}
\rightline{SUNY BING 7/1/98}
\rightline{hep-th/9807047}
\vspace{1mm}
\begin{center}
{\bf  DIRAC'S CONTOUR REPRESENTATION FOR PARAPARTICLES}\\
\vspace{2mm}
Sicong Jing\footnote{On leave from:  Department of
Modern Physics, University of Science and Technology of 
China, Hefei, 230026, P.R. China.} and 
Charles A. Nelson\footnote{Electronic address: cnelson @ 
bingvmb.cc.binghamton.edu }\\
{\it Department of Physics, State University of New York at 
Binghamton\\
Binghamton, N.Y. 13902-6016}\\[2mm]
\end{center}


\begin{abstract} 
Dirac's contour representation is extended to parabose and 
parafermi systems by use of deformed algebra techniques.  In 
this analytic representation the action of the paraparticle 
annihilation operator is equivalent to a deformed 
differentiation which encodes the statistics of the 
paraparticle. In the parafermi case, the derivative's 
ket-domain is degree $p$ polynomials.
\end{abstract}

\end{titlepage}

\section{Introduction}

Recently for boson systems, there has been interest in the 
Dirac contour representation [1-3] because analytic 
representations are frequently important in the 
analysis of quantum field theoretic systems. The contour 
aspect of this representation is based on Heitler's contour 
integral form of the $\delta$-function [4].  This   
representation provides a natural enlargement of the usual 
Hilbert space description because the expansion coefficients 
are distributions, instead of functions.

In this paper, we extend the representation to parabose  
and parafermi systems [5-6] by using deformed algebra 
techniques [7-9].  In formal appearance, the resulting 
contour representations are similar in the two cases.  The 
non-trivial differences arise in the appearance of deformed 
brackets, $[n]$, and a deformed differentiation, 
$\frac {D}{D z}$, in the parabose case, versus ``curly 
braces"
$\{ n \}$ and $\frac{ {\cal D} }{ {\cal D} z}$ in the 
parafermi case.
When $p=1$, the parafermi extension gives a contour 
representation for ordinary fermions.  

In Sec. 2, the Dirac contour representation for parabose 
systems is constructed and its properties are studied.  Sec. 
3 is devoted to the extension to parafermi systems.    
Finally, in Sec. 4 there is a brief summary discussion.
 
\section{Dirac's Contour Representation for Parabosons}

The trilinear commutation relations in the single-mode 
parabose case are
\beq
[a,\{a^{\dagger },a\}]=2a,\; \; [a,(a^{\dagger 
})^2]=2a^{\dagger}. 
\eeq
where $a^{\dagger}$, $a$ are the parabose creation and 
annihilation operators.  Actually, these relations can be 
replaced [8] by deformed bilinear ones
\beq
[a,a^{\dagger }]=1+(p-1)(-)^{N_B},\; \; \{a,a^{\dagger } 
\}=p+2 N_B
\eeq
where $N_B$ is the parabose number operator and $p$
is the order of the parastatistics.
The number basis for the single-mode parabose system is 
$$
|n> = \frac{(a^{\dagger })^n}{%
\sqrt{[n]!} },
$$
with
\beq
[n]=n+\frac{p-1}2\left( 1-(-)^n\right) ,\; [n]!=[n][n-
1]\cdots
[1],\;[0]! \equiv 1. 
\eeq
Thus the eigenbras\footnote{Note that the dual space $<n|$ is 
not obtained by Hermitian
conjugation, see [1-3].} and eigenkets of $N_B$ are 
respectively represented by 
\begin{equation}
<n|\rightarrow \frac{\sqrt{[n]!}}{z^{n+1}},\; \; 
|m>\rightarrow 
\frac{z^m}{%
\sqrt{[m]!}}
\end{equation}
in the 
Dirac contour representation associated with the parabose 
Hilbert space ${\cal H}_B$. In this representation, the 
``bra-ket'' inner product is 
defined by 
\begin{equation}
<n|m>\rightarrow \oint_C\;\frac{dz}{2\pi 
i}\;\sqrt{\frac{[m]!}{[n]!}}%
\frac{z^n}{z^{m+1}}\;=\sqrt{\frac{[m]!}{[n]!}}\frac 
1{m!}\frac{d^m}{dz^m}%
z^n|_{z=0}\;=\delta _{mn}
\end{equation}
where $C$ is an counterclockwise contour enclosing the origin 
of 
the complex $z$
plane. Arbitrary bra and ket states in ${\cal H}_B$, which 
can 
be 
expanded as $%
<f|= \sum_{n=0}^\infty f_n^{*}<n|,\;|f> = 
\sum_{m=0}^\infty f_m|m>,$  are represented by 
\begin{equation}
<f|\rightarrow \sum_{n=0}^\infty 
f_n^{*}\frac{\sqrt{[n]!}}{z^{n+1}}%
=f_b(z),\;|f>\rightarrow \sum_{m=0}^\infty 
f_m\frac{z^m}{\sqrt{[m]!}}%
=f_k(z)
\end{equation}
where the subscripts $b (k)$ refer to bra (ket).
When $\sum |f_n|^2=1$, the normalized function $f_k(z)$ is an 
analytic 
function in the complex plane $z\;\epsilon \;C$ . If $C$ is 
enlarged to
include the point at infinity \newline ( ${\bar C} =C+\infty 
$\ ), the 
function $f_b(z)$
is also an analytic function in the neighborhood of the 
infinity point in ${\bar C}$ . From (5) the inner product of 
two such 
states 
is 
\begin{equation}
<f|g>\rightarrow \oint_C\;\frac{dz}{2\pi 
i}\;f_b(z)\;g_k(z)\;=\sum_{n=0}^%
\infty f_n^{*}\;g_n
\end{equation}

An arbitrary operator $\Theta $ in  ${\cal H}_B$ with $\Theta
=\sum_{m,n=0}^\infty \Theta _{mn}|m><n|,\;\Theta 
_{mn}=<m|\Theta |n>,$ has
the Dirac contour representation
\begin{equation}
\Theta \rightarrow \Theta (z,w)=\sum_{m,n=0}^\infty \Theta 
_{mn}\sqrt{\frac{%
[n]!}{[m]!}}\frac{z^m}{w^{n+1}}.
\end{equation}
So,
\begin{equation}
\Theta |g>\rightarrow \oint_B\;\frac{dw}{2\pi i}\Theta
(z,w)g_k(w),\; <f|\Theta \rightarrow \oint_B\;\frac{dw}{2\pi 
i}f_b(w)\Theta
(w,z)
\end{equation}
represent new states where the counterclockwise contour $B$ 
encloses the 
origin of the
complex $w$ plane. The product of two operators $\Theta _1$, 
$\Theta _2$ in $%
{\cal H}_B$ takes the form of a generalized convolution 
\begin{equation}
\Theta _1\Theta _2\rightarrow \oint_B\;\frac{dv}{2\pi 
i}\Theta _1 (z,v)\Theta _2
(v,w) 
\end{equation}

In the parabose number representation, $a^{\dagger 
}|n>=\sqrt{[n+1]}|n+1>$%
, which gives $(a^{\dagger })_{mn}=\sqrt{[n+1]}\delta 
_{m,n+1}$, e.g. see [6].
So from (8)%
$$
a^{\dagger }\rightarrow \sum_{m,n=0}^\infty 
\sqrt{[n+1]}\delta _{m,n+1}%
\sqrt{\frac{[n]!}{[m]!}}\frac{z^m}{w^{n+1}}=
\sum_{n=0}^\infty  
\frac{z^{n+1}}{w^{n+1}}, 
$$
which converges to $z(w-z)^{-1}$ when $|w|>|z|$. For 
$|z|>|w|$ , the sum
diverges.  However, then the point $z$ in first equation in 
(9) lies outside 
the contour $B$
and so (9) gives zero since $g_k(w)$ is an arbitrary 
analytic  
function. So in 
this contour representation\footnote{Here $%
\theta (x)=1$ for $ x>0$ and $\theta (x)=0$ for $ x<0$.}, we 
write 
\begin{equation}
a^{\dagger }\rightarrow \frac z{w-z}\;\theta (|w|-|z|),
\end{equation}
Also note, by (9),%
$$
a^{\dagger }|n>\rightarrow \oint_B\;\frac{dw}{2\pi i}\;\frac 
z{w-z}\;\frac{%
w^n}{\sqrt{[n]!}}=\frac{z^{n+1}}{\sqrt{[n]!}}=\sqrt{[n+1
]}\frac{z^{n+1}%
}{\sqrt{[n+1]!}} 
$$
which agrees with $a^{\dagger }|n>=\sqrt{[n+1]}|n+1>$, and 
\begin{equation}
<0|a^{\dagger }\rightarrow \oint_C\;\frac{dz}{2\pi i}\;\frac 
1z\frac
z{w-z}=0,\;\;(|w|>|z|)
\end{equation}
so $a^{\dagger }$ does annihilate the bra state 
$<0|\rightarrow \frac
1z$.

In the number representation, $a|n>=\sqrt{[n]}|n-1>$, so 
$a_{mn}=\sqrt{%
[n]} \; \delta _{m,n-1}$ and%
$$
a\rightarrow \sum_{m,n=0}^\infty \sqrt{[n]}\delta _{m,n-
1}\sqrt{\frac{%
[n]!}{[m]!}}\frac{z^m}{w^{n+1}}=\frac 
1{w^2}\sum_{n=0}^\infty
[n]\left( \frac zw\right) ^{n-1}, 
$$
\begin{equation}
=\left( \frac 1{(w-z)^2}+\frac{p-1}{2z}\frac 1{w-z}-\frac{p-
1}{2z}\frac
1{w+z}\right) \theta (|w|-|z|).
\end{equation}
Obviously, (13) reduces to Dirac's boson result [1] when 
$p=1$.
With the definition [9] of deformed differentiation (c.f. 
Eq.(3) above)
\begin{equation}
\frac {D}{D z}f(z) \equiv \frac d{dz}f(z)+\frac{p-
1}{2z}\left( 
f(z)-f(-
z)\right) ,
\end{equation}
we can write (13) as
\begin{equation}
a\rightarrow (\frac D{Dz}\frac 1{w-z})\;\theta (|w|-|z|),
\end{equation}
Again, from (9) we obtain 
$$
a|n>\rightarrow \oint_B\;
\frac{dw}{2\pi i}\;\left( \frac 1{(w-z)^2}+\frac{p-
1}{2z}\frac 1{w-z}-\frac{%
p-1}{2z}\frac 1{w+z}\right) \frac{w^n}{\sqrt{[n]!}} 
$$
\beq 
=\left( n+\frac{p-1}%
2-\frac{p-1}2(-)^n\right) \frac{z^{n-
1}}{\sqrt{[n]!}}\,=\sqrt{[n]}\frac{%
z^{n-1}}{\sqrt{[n-1]!}}
\eeq
which agrees with $a|n>=\sqrt{[n]}|n-1>$.

Then from the convolution formula (10), we obtain%
$$
a^{\dagger }a\rightarrow \oint_C\;\frac{dv}{2\pi 
i}\;\frac{z}{v-z}\left(
\frac 1{(w-v)^2}+\frac{p-1}{2v(w-v)}-\frac{p-
1}{2v(w+v)}\right)  
$$
\begin{equation}
=\left( \frac{z}{(w-z)^2}+\frac{p-1}{2(w-z)}-\frac{p-
1}{2(w+z)}%
\right) \theta (|w|-|z|)
\end{equation}
and 
$$
aa^{\dagger }\rightarrow \oint_C\;\frac{dv}{2\pi i}\;\left( 
\frac
1{(v-z)^2}+\frac{p-1}{2z(v-z)}-\frac{p-
1}{2z(v+z)}\right) \;\frac
v{w-v} 
$$
\begin{equation}
=\left( \frac{w}{(w-z)^2}+\frac{p-1}{2(w-z)}+\frac{p-
1}{2(w+z)}%
\right) \theta (|w|-|z|).
\end{equation}
In the contour integration in (18), the points $z$ and 
$\left( 
-z\right) $
both lie inside the contour $C$. From (17-18),%
$$
(N_B +\frac p2)\;|n>=\frac 12(a^{\dagger }a+aa^{\dagger 
})\;|n> 
$$
\begin{equation}
\rightarrow \oint_B\;\frac{dw}{2\pi i}\;\frac 12\left( 
\frac{z+w}{(w-z)^2}+%
\frac{p-1}{w-z}\right) 
\;\frac{w^n}{\sqrt{[n]!}}\;=\;(n+\frac p2)\frac{z^n%
}{\sqrt{[n]!}},
\end{equation}
$$
<n|\;(N_B +\frac p2)=(n|\;\frac 12(a^{\dagger }a+aa^{\dagger 
}) 
$$
$$
\rightarrow \oint_B\;
\frac{dw}{2\pi i}\;\frac{\sqrt{[n]!}}{2w^{n+1}}\;\frac 
12\left( \frac{z+w}{%
(z-w)^2}+\frac{p-1}{z-w}\right) 
\;=\frac{\sqrt{[n]!}}{2n!}\frac{d^n}{dw^n}%
\left( \frac{z+w}{(z-w)^2}+\frac{p-1}{z-w}\right) |_{w=0} 
$$ 
\begin{equation}
=\;(n+\frac p2)%
\frac{\sqrt{[n]!}}{2z^{n+1}},
\end{equation}
which explicitly verifies that $|n>$ and $<n|$ are 
eigenstates of the
parabose number operator $N_B$.

Similarly, for integer powers of $a$ and $a^{\dagger }$%
\begin{equation}
\begin{array}{c}
(a^{\dagger })^m\rightarrow 
\frac{z^m}{w-z}\;\theta (|w|-|z|), \\ (a)^n\rightarrow (\frac 
{D^n}{Dz^n}\frac
1{w-z})\;\theta (|w|-|z|).
\end{array}
\end{equation}
So from convolution formula, the representations for their 
normal-ordered
and antinormal-ordered products are 
\begin{equation}
\begin{array}{c}
(a^{\dagger })^ma^n\rightarrow \oint_C\;
\frac{dv}{2\pi i}\;\frac{(z)^m}{v-z}\;\left( 
\frac{D^n}{Dv^n}\;\frac
1{w-v}\right) =z^m(\frac{D^n}{Dz^n}\;\frac 1{w-
z})\theta
(|w|-|z|), \\ (a)^n(a^{\dagger })^m\rightarrow \oint_C\;
\frac{dv}{2\pi i}\;\left( \frac{D^n}{Dz^n}\;\frac 1{v-
z}\right) \;\frac{%
v^m}{w-v}=\frac{D^n}{Dz^n}\oint_C\;\frac{dv}{2\pi 
i}\;\left( \frac
1{v-z}\right) \;\frac{v^m}{w-v} \\ 
=\frac{D^n}{Dz^n}(\frac{z^m}{%
w-z})\theta (|w|-|z|)
\end{array}
\end{equation}
Furthermore, for an arbitrary ket $|g>$ in ${\cal H}_B $ , 
\begin{equation}
\begin{array}{c}
(a^{\dagger })^ma^n|g>\rightarrow \oint_B\;
\frac{dw}{2\pi i}\;z^m\;\left( \frac{D^n}{Dz^n}\;\frac 1{w-
z}\right) g_k(w) \\ =z^m
\frac{D^n}{Dz^n}\oint_B\;\frac{dw}{2\pi i}\;\left( \frac 
1{w-z}\right)
g_k(w)=z^m\frac{D^n}{Dz^n}g_k(z), \\ (a)^n(a^{\dagger 
})^m|g>\rightarrow
\oint_B\;
\frac{dw}{2\pi i}\;\left( \frac{D^n}{Dz^n}\;\frac{z^m}{w-
z}\right) \;g_k(w)
\\ =\frac{D^n}{Dz^n}\left( z^m\oint_B\;\frac{dw}{2\pi 
i}\;\left( \frac
1{w-z}\right) g_k(w)\right) =\frac{D^n}{Dz^n}\left( 
z^mg_k(z)\right) ,
\end{array}
\end{equation}
which shows that in the Dirac contour representation (as well 
as in the
Bargmann representation [10]) the action of the parabose 
operators 
$a^{\dagger }$
and $a$ is respectively equivalent\footnote{An instructive 
exercise is to verify the trilinear commutation relations;
note in general, $\frac D{Dz}(fg)\neq (\frac D{Dz}f)g+f(\frac 
D{Dz}g).$ } to multiplication by $z$ 
and deformed
differentiation, $\frac{ D }{Dz}$ , of the analytic
function 
$g_k(z)$.

As the last topic in this section, we give the Dirac contour 
representation
for the unnormalized parabose coherent state $|\alpha >$:
\begin{equation}
|\alpha >=\sum_{n=0}^\infty \frac{\alpha 
^n}{\sqrt{[n]!}}|n>\;\rightarrow
\sum_{n=0}^\infty \frac{\alpha 
^n}{\sqrt{[n]!}}\;\frac{z^n}{\sqrt{[n]!}}%
=E(\alpha z)
\end{equation}
where $E(z)\equiv \sum_{n=0}^\infty 
\frac{z^n}{\sqrt{[n]!}}$ is an entire
function. By (9), 
\begin{equation}
a\;|\alpha >\rightarrow \oint_B\;\frac{dw}{2\pi i}\;\left( 
\frac
D{Dz}\;\frac 1{w-z}\right) \;E(\alpha w)=\alpha E(\alpha z),
\end{equation}
which verifies $a\;|\alpha >=\alpha \;|\alpha >$. By (4), the 
corresponding
bra state $<\alpha |$\ in the Dirac contour representation is 
\begin{equation}
<\alpha |=\sum_{n=0}^\infty <n|\frac{(\alpha 
^{*})^n}{\sqrt{[n]!}}%
\;\rightarrow \sum_{n=0}^\infty 
\frac{\sqrt{[n]!}}{z^{n+1}}\frac{(\alpha
^{*})^n}{\sqrt{[n]!}}=\frac 1{z-\alpha ^{*}}, \; \; 
( |z| > |\alpha|).
\end{equation}
So by (9),
\begin{equation}
<\alpha |a^{\dagger }\rightarrow \oint_B\;\frac{dw}{2\pi 
i}\;\left( \frac
1{w-\alpha ^{*}}\right) \;\frac w{z-w}=\frac{\alpha ^{*}}{z-
\alpha ^{*}}
 \; \;  ( |z| > |\alpha|)
\end{equation}
which shows that $ 
<\alpha |a^{\dagger }=<\alpha |\;\alpha ^{*} $.  
This means that in the Dirac contour representation, the dual 
vectors $<\alpha |$ of the parabose coherent states 
are the eigenbra vectors of the parabose creation operator 
$a^{\dagger }$ with eigenvalue $\alpha$.

The contour form of the resolution of unity [1,2]  
associated with  $<\alpha|$  and the $a^{\dagger}$ 
eigenvectors $|z>^{'}$ is eq.(13) in [12].

\section{Dirac's Contour Representation for Parafermions}

The parafermi trilinear commutation relation in the single-
mode case is
\begin{equation}
[f,[f^{\dagger },f]]=2f
\end{equation}
where $f^{\dagger },f$ are the parfermi creation and 
annihilation operators.
This trilinear relation can also be rewritten in bilinear 
form in terms of
deformed oscillators[7]
\begin{equation}
\{f,f^{\dagger }\}=p+2pN_f-2N_f^2,\;[f,f^{\dagger }]=p-2N_f
\end{equation}
where $N_f$ is the parafermi number operator. We also define 
a deformed
parafermi number operator 
\begin{equation}
\{N_f\}=N_f(p+1-N_f)
\end{equation}

To obtain the ordinary fermions for $p=1$, we set 
$N_f=\{N_f\}$ which
implies $N_f^2=N_f$ and then the bilinear relations (29) 
imply $%
N_f=f^{\dagger }f$ and the ordinary fermi algebra 
$\{f,f^{\dagger }\}=1$.
The eigenstates of $N_f$ are
\begin{equation}
|n>=\frac{(f^{\dagger })^n}{\sqrt{\{n\}!}}|0>,\;N_f|n>=n|n>,
\end{equation}
with  
\begin{equation}
\{n\}=n(p+1-n),\;\;\{n\}!=\{n\}\{n-1\}\cdots 
\{1\},\;\;\{0\}!=1
\end{equation}
From (32), $\{0\}=\{p+1\}=0$ and $\{n\}<0$ for $n>p+1$. So 
for parafermi systems, we consider $n$ as non-negative 
integers in the range $%
0\leq n\leq p$. In this number basis, we also have 
\begin{equation}
f|n>=\sqrt{\{n\}}|n-1>,\;\;f^{\dagger 
}|n>=\sqrt{\{n+1\}}|n+1>\;\;(0\leq
n\leq p)
\end{equation}

In order to construct the contour representation for 
parafermi systems, we
introduce a polynomial basis[7]. Let ${\cal H}$ be the set 
of analytic
functions 
$$
f(z)=\sum_{n=0}^\infty c_nz^n, 
$$
in the complex plane $C$ and ${\cal J}_p$ be the projection 
operator which
projects the function $f(z)$ to the polynomial ${\cal 
J}_pf(z)$ of degree $p$
\begin{equation}
{\cal J}_pf(z)=\sum_{n=0}^pc_nz^n\;\epsilon \;{\cal J}_p{\cal 
H.}
\end{equation}
Thereby, the space spanned by the parafermi number basis 
$|n>$ is
identified with the polynomial space ${\cal J}_p{\cal H}$ 
spanned by the 
basis $\frac{%
z^n}{\sqrt{\{n\}!}}$,
\begin{equation}
|n>\rightarrow \frac{z^n}{\sqrt{\{n\}!}},\;(n=0,1,\cdots p).
\end{equation}
We denote the dual space of ${\cal H}$ by ${\cal H}^{*}$, 
which is the set
of functions 
$$
g(z)=\sum_{n=0}^\infty \frac{d_n}{z^{n+1}}. 
$$
When $\sum |d_n|^2=1$, these functions are also analytic in 
the
neighborhood of the infinity point in $\bar C$. In  ${\cal 
H}^{*}$ we
similarly introduce a projection operator ${\cal J}_p^{*}$ 
which projects
the function $g(z)$ to a truncated one ${\cal J}_p^{*}g(z)$ 
of minus degree $%
p+1$%
\begin{equation}
{\cal J}_p^{*}g(z)=\sum_{n=0}^p\frac{d_n}{z^{n+1}}\epsilon 
\;{\cal J}_p^{*}%
{\cal H}^{*}.
\end{equation}
Thus, the space spanned by the dual vectors $<n|$ of the 
parafermi number
basis $|n>$ is identified with the space ${\cal J}_p^{*}{\cal 
H}^{*}$ spanned by
\begin{equation}
<n|\rightarrow 
\frac{\sqrt{\{n\}!}}{z^{n+1}},\;\;(n=0,1,\cdots p).
\end{equation}
In this representation, we define the inner product of $<n|$ 
and $|m>$
by
\begin{equation}
<n|m>\rightarrow \oint_C\frac{dz}{2\pi \iota 
}\sqrt{\frac{\{n\}!}{\{m\}!}}%
\frac{z^m}{z^{n+1}}=\delta _{mn}
\end{equation}
where the contour $C$ is defined as in (5).

Again guided by (30) with $\{N\}\rightarrow z\frac { {\cal D}  
}{ {\cal D} z}$ and 
$N\rightarrow
z\frac d{dz},$ in the parafermi case we define the deformed 
derivative
\begin{equation}
\frac{ {\cal D} }{ {\cal D} z}f(z)\equiv \frac d{dz}(p+1-
z\frac d{dz})
\end{equation}
which gives for instance 
\begin{equation}
\frac{ {\cal D} }{ {\cal D} z}z^n=\{n\}\;z^{n-
1},\;\;(n=0,1,\cdots p).
\end{equation}
The multiplication of the function ${\cal J}_pf(z)$ by $z$ 
can be regarded
as a map from the subspace ${\cal J}_p{\cal H}$ into the 
subspace ${\cal J}_p%
{\cal H-J}_0{\cal H}$ since
\begin{equation}
 z \; \sum_{n=0}^pc_nz^n=\sum_{n=1}^pc_{n-1}z^n\;\epsilon 
\;({\cal J}_p%
{\cal H}-{\cal J}_0{\cal H).}
\end{equation}
Also the derivative $\frac d{dz}$ is a map from  ${\cal 
J}_p{\cal H}$ into $%
{\cal J}_{p-1}{\cal H}$ since
\begin{equation}
\frac d{dz}\sum_{n=0}^pc_nz^n=\sum_{n=0}^{p-
1}(n+1)\;c_{n+1}\;z^n\;\epsilon
\;{\cal J}_{p-1}{\cal H.}
\end{equation}
Similarly, multiplication of ${\cal J}_p^{*}g(z)$ by $z$ is a 
map from $%
{\cal J}_p^{*}{\cal H}^{*}$ into ${\cal J}_{p-1}^{*}{\cal 
H}^{*}+{\cal J}_0%
{\cal H,}$%
\begin{equation}
 z \; \sum_{n=0}^p\frac{d_n}{z^{n+1}}=d_0+\sum_{n=0}^{p-
1}\frac{d_{n+1}}{%
z^{n+1}}\;\epsilon \;({\cal J}_{p-1}^{*}{\cal H}^{*}+.{\cal 
J}_0{\cal H),}
\end{equation}
and the derivative $\frac d{dz}$ is a map from ${\cal 
J}_p^{*}{\cal H}^{*}$
into ${\cal J}_p^{*}{\cal H}^{*}-{\cal J}_0^{*}{\cal 
H}^{*}{\cal ,}$%
\begin{equation}
\frac d{dz}\sum_{n=0}^p\frac{d_n}{z^{n+1}}=-
\sum_{n=1}^p\;\frac{n\;d_{n-1}}{%
z^{n+1}}\;\epsilon \;({\cal J}_p^{*}{\cal H}^{*}-{\cal 
J}_0^{*}{\cal H}^{*}).
\end{equation}
Note that when $p=1$, parafermi deformed differentiation 
reduces to 
ordinary differentiation in the subspace ${\cal J}_p%
{\cal H}$.

For simplicity, we shall omit the explicit projection symbols 
${\cal J}_p$
and ${\cal J}_p^{*}$ in the following so in the parafermi 
case it is to be
remembered that the domains are the subspaces 
${\cal J}_p{\cal H}$
and ${\cal J}_p^{*}{\cal H}^{*}$. With this understanding, 
the formalism is
analogous to that for parabose systems: Arbitrary bra and ket 
states in the
parafermi Hilbert space ${\cal H}_f$ which can be expanded as 
$%
<g|=\sum_{n=0}^pg_n^{*} <n|,\;\;|g>=\sum_{m=0}^pg_m |m>$, are 
represented by
\begin{equation}
<g|\rightarrow 
\sum_{n=0}^pg_n^{*}\frac{\sqrt{\{n\}!}}{z^{n+1}}%
=g_b^{(f)}(z),\;\;|g>\;\rightarrow 
\sum_{m=0}^pg_m\frac{z^m}{\sqrt{\{m\}!}}%
=g_k^{(f)}(z)
\end{equation}
and the inner product of two such vectors is
\begin{equation}
<f|g>\rightarrow \oint_C\frac{dz}{2\pi \iota 
}f_b^{(f)}(z)\;g_k^{(f)}(z)=%
\sum_{n=0}^pf_n^{*}\;g_n.
\end{equation}
For an operator $A$ defined on ${\cal H}_f$ by 
$A=\sum_{m=0}^pA_{mn}|m><n|$,
there is the contour representation
\begin{equation}
A\rightarrow 
A(z,w)=\sum_{m,n=0}^pA_{mn}\sqrt{\frac{\{n\}!}{\{m\}!}}\frac{
z^m%
}{z^{n+1}}.
\end{equation}
so Eqs.(9-10) again follow. As in the parabose case, we find 
that 
$$
f^{\dagger }\rightarrow \frac z{w-z}\;\;\theta (|w|-|z|), 
$$
$$
f\rightarrow (\frac{ {\cal D} }{ {\cal D} z}\frac 1{w-
z})\;\;\theta (|w|-
|z|), 
$$
$$
f^{\dagger }f\rightarrow z\;(\frac{ {\cal D} }
{ {\cal D} z}\frac 1{w-z})\;\;\theta
(|w|-|z|), 
$$
\begin{equation}
f\;f^{\dagger }\rightarrow (\frac{ {\cal D} }
{ {\cal D} z}\frac z{w-z})\;\;\theta
(|w|-|z|).
\end{equation}
Also, the unnormalized parafermi coherent state $|\beta >_f$ 
is represented
by
\begin{equation}
|\beta >_f=\sum_{n=0}^p\frac{\beta 
^m}{\sqrt{\{n\}!}}|n>\;\rightarrow
\sum_{n=0}^p\frac{\beta ^m}{\{n\}!}\;\equiv \;E_f(\beta z)
\end{equation}
where $E_f(z)$ is a polynomial. By (9), $f|\beta >_f=\beta 
\;|\beta >_f.$
Furthermore, by (45) the corresponding bra state in this 
contour
representation is 
\begin{equation}
_f<\beta |=\sum_{n=0}^p\frac{(\beta 
^{*})^m}{\sqrt{\{n\}!}}<n|\;\rightarrow
\sum_{n=0}^p\frac{(\beta ^{*})^m}{z^{n+1}}=\frac 1{z-\beta
^{*}},\;\;(|z|>|\beta |),
\end{equation}
which satisfies $_f<\beta |\;f^{\dagger }=_f<\beta |\;\beta 
^{*}$.  The contour form of the resolution of unity  
associated with $_f<\beta |$ is 
formally the same as that for parabosons, see eq.(13) in 
[12], but it is 
restricted to the parafermi subspace.

\section{Discussion}

In summary, in this paper we extend the Dirac contour 
representation to the single-mode parabose and parafermi 
systems.  In these extensions we use deformed bilinear 
commutation relations to replace the intrinsic trilinear 
relations of paraparticles.  Thus the formalism is similar to 
that for ordinary bose systems, the non-trivial differences 
include (i) the appearance of $\sqrt{[n]!}$ or 
$\sqrt{\{n\}!}$ in parasystems, in place of $\sqrt{n!}$, in 
expressions which represent states and operators versus the 
number basis, and (ii) the representation of the action of 
the paraparticle annihilation operator by a deformed 
differentiation $\frac {D}{D z}$ or 
$\frac{ {\cal D} }{ {\cal D} z}$, instead of by ordinary 
differentiation. In a similar manner, the Dirac contour 
representation can also be extended to many other deformed 
oscillator systems such as the q-deformed harmonic 
oscillator, the Arik-Coon oscillator, and the q-deformed 
parabose and parafermi oscillators[11].  The 
action of the oscillator's annihilation operator is 
equivalent to a deformed differentiation which encodes the   
statistical properties of the oscillator.  
As in the case of parafermions, see (39-44), the definition 
of the deformed derivative always must include specification 
of its domain. Deformed integration, i.e. the inverse 
operation, can therefore also be defined. 

This work was partially supported by the National Natural 
Science Foundation of China and by U.S. Dept. of Energy 
Contract No. DE-FG 02-96ER40291.

\end{document}